\documentstyle[11pt,epsfig]{article}
\newcommand{\be}{\begin{eqnarray}}
\newcommand{\ee}{\end{eqnarray}}
\newcommand{\bra}[1]{\mbox{$\langle\, #1 \mid$}}

\newcommand{\ket}[1]{\mbox{$\mid #1\,\rangle$}}

\newcommand{\pro}[2]{\mbox{$\langle\, #1 \mid #2\,\rangle$}}
\newcommand{\expec}[1]{\mbox{$\langle\, #1\,\rangle$}}
\newcommand{\expecl}[1]{\mbox{$\left\langle\, 
	    \strut\displaystyle{#1}\,\right\rangle$}}

\renewcommand{\a}{\hat a}
\newcommand{\ac}{\hat a^{\dagger}}
\renewcommand{\b}{\hat b}
\newcommand{\bc}{\hat b^\dagger}
\title{Gravitational Collapse of a Shell of Quantized Matter}
\author{G.L. Alberghi\thanks{e-mail: alberghi@bo.infn.it},\ 
R. Casadio\thanks{e-mail: casadio@bo.infn.it}, 
\ G.P. Vacca\thanks{e-mail: vacca@bo.infn.it} 
\ and G. Venturi\thanks{e-mail: armitage@bo.infn.it}\\
 \\
{\em Dipartimento di Fisica, Universit\`a di Bologna} \\
{\em and} \\
{\em Istituto Nazionale di Fisica Nucleare, 
Sezione di Bologna, Italy}}
\begin{document}
%
%
\maketitle
\begin{abstract}
The semi-classical collapse, including lowest order back-reaction,
of a thin shell of self-gravitating quantized matter is illustrated.
The conditions for which self-gravitating matter forms a thin shell
are first discussed and an effective Lagrangian for such matter
is obtained.
The matter-gravity system is then quantized, the semi-classical limit
for gravitation is taken and the method of adiabatic invariants is
applied to the resulting time dependent matter Hamiltonian.
The governing equations are integrated numerically,
for suitable initial conditions, in order to illustrate the effect of
back-reaction, due to the creation of matter, in slowing down the
collapse near the horizon.
\end{abstract}
%
\pagestyle{plain}
\raggedbottom
\setcounter{page}{1}
\section{Introduction}
\setcounter{equation}{0}
Much effort has been dedicated to the classical dynamics of collapsing
gravitational bodies (see {\em e.g.} Refs.~\cite{tolman,oppenheimer,Israel}
for early works and \cite{singh} for a summary of more recent developments).
In particular, the collapse of an isotropic homogeneous sphere of
dust has been studied using the Arnowitt, Deser and Misner (ADM)
construction \cite{adm} both on relating the dust associated scalar 
field to time \cite{lund} and including the quantum nature of the collapsing 
matter through a Born-Oppenheimer approach to the
Wheeler-DeWitt (WDW) equation describing the coupled matter-gravity 
system (see \cite{brout,bv,bertoni} and Refs. therein).
Within the latter approach the evolution of quantum matter (dust) was
studied both in the adiabatic approximation \cite{cv} and on relaxing it 
\cite{cfv}.
\par
Homogeneous dust in flat space-time without boundaries corresponds 
to a fluid with (possibly time-dependent) density and zero pressure 
and can be related to the homogeneous mode of a free massive real 
scalar field.
When dust self-gravitates and is confined into a sphere of finite 
volume, the metric inside the sphere is a section of a Friedmann 
universe (in the language of cosmological models) with constant 
spatial curvature and the radius of the sphere depends on time 
\cite{oppenheimer,stephani}.
As we have previously indicated \cite{cv,cfv}, using the homogeneous
mode of a scalar field to describe a sphere of dust is still plausible 
as long as one considers spheres whose radius is much greater than the 
scalar field Compton wave-length. 
In fact for such a case edge effects are negligible \cite{moss}.
\par
We have previously, through the use of adiabatic invariants 
\cite{lewis-ries} for time dependent Hamiltonians, relaxed the adiabatic
approximation and thus allowed for the creation of matter (dust
particles).
However one will then eventually obtain a non-zero pressure
and have deviations from the dust approximation.
\par
Rather than study an isotropic, but inhomogeneous, system it is easier
to address the collapse of a gravitating matter shell
which we now briefly review, also in order to introduce our notation.
\par
In general the dynamics of a system consisting of self-gravitating matter
in a space-time manifold of 4-volume $V$ with boundary $B$ can 
be derived from the Einstein-Hilbert action with a matter term
\be
S=\int _V d^4x\,\sqrt{-g}\,\left[{{\cal R}^{(4)}\over16\,\pi\,G}
+{\cal L}_M\right]+
{1\over{8\,\pi\,G}}\,\int _B d^3x\,\sqrt{-\tilde g}\,{\cal K}
=\int L\, d \tau
\ ,
\ee
where $G$ is Newton's constant, $g$ the determinant of the 
4-dimensional metric, ${\cal R}^{(4)}$ the Ricci scalar, ${\cal L}_M$ 
the Lagrangian density for matter and ${\cal K} $ the trace 
of the extrinsic curvature of the boundary $B$ on which a 3-dimensional 
metric with determinant $\tilde g$ is induced.
\par
When matter is localized on an isotropic shell, we can introduce
an adapted radial coordinate $r$ such that the shell is parametrized
by $r_{in}\le r\le r_{out}$. 
Then the external geometry is uniquely determined by Birkhoff's 
theorem to be given by \cite{asy}
\be 
ds^2 _{in\,(out)}=-A_{in\,(out)}\,dt^2+A_{in\,(out)}^{-1}\,dr^2
+r^2\,d\Omega^2
\ , 
\label{metric}
\ee 
where $A_{in}=1$ for $r<r_{in}$, corresponding to a Minkowski metric 
inside the shell, and $A_{out}=1-2\,G\,M/r$ for $r>r_{out}$,
corresponding to a Schwarzschild metric of ADM mass $M$ outside the shell.  
\par
If the Schwarzschild radius of the shell, $R_H=2\,G\,M$, is much greater 
than the thickness $d\sim r_{out}-r_{in}$ of the shell (as we shall show 
in the next section is true in our case) one can consider the ``thin limit'' 
$d/r_{in}\to 0$ (with fixed matter proper energy), define 
$R\equiv (r_{in}+r_{out})/2$ 
and the Lagrangian assumes the form \cite{Balbinot,Hajicek}
\be
L&=&-\left[{{R\,\dot R}\over G}\,
\tanh ^{-1}\left({{\dot R}\over\beta}\right)\right]^{in}_{out}
+{R\over G}\,(\beta_{in}-\beta_{out})-m 
\nonumber \\
&=&-{{R\,\dot R}\over G}\left[
\tanh^{-1}\left({{\dot R}\over{\sqrt{1+\dot R^2}}}\right)
-\tanh^{-1}\left({{\dot R}\over{\sqrt{1-{2\,G\,M\over R}+\dot R^2}}}\right)  
\right]+
\nonumber \\
&&+{R\over G}\,\left(\sqrt{1+\dot R^2}-\sqrt{1-{2\,G\,M\over R}+\dot R^2}
\right)-m 
\ ,
\label{gravilag}
\ee
where $\beta_{in\,(out)}=\sqrt{A_{in\,(out)}+\dot R^2}$,
$[a]^{in}_{out}\equiv a(r_{in})-a(r_{out})$ for any function $a(r)$ 
and a dot denotes the derivative with respect to the proper time $\tau$ 
($d\tau=A^{1/2}\,dt$).
The quantity $m$ is the volume integral of the matter Hamiltonian 
density and is equal to the proper energy of matter.
\par
The next step is the construction of the Hamiltonian, 
\be 
H&=&P_R\,\dot R-L=-{R\over G}\,(\beta_{in}-\beta _{out})+m
\nonumber \\
&=&m-{{\sqrt{2}\,R}\over G}\,\left[
1-{G\,M\over R}-\sqrt{1-{2\,G\,M\over R}}\,
\cosh\left({G\,P_R\over R}\right)\right]^{1/2}
\ ,
\label{hamiltonian}
\ee
with
\be
P_R\equiv {\partial L\over\partial \dot R}= 
-{R\over G}\,\left[
\tanh^{-1}\left({\dot R\over\beta}\right)\right]^{in}_{out}
\ .
\ee
It is a general feature of gravitational minisuperspace systems
that the equation of motion is obtained on imposing the constraint 
$H=0$, which expresses the invariance of the action under a time
reparametrization.
In our case the equation of motion takes the form \cite{Israel}
\be
\sqrt{1+\dot R^2}-\sqrt{1-{2\,G\,M\over R}+\dot R^2 }={G\,m\over R}
\  .
\ee
By squaring twice this expression becomes
\be 
\dot R^2-{G\,M\over R}-{G^2\,m^2 \over4\,R^2}={M^2\over m^2}-1
\ .
\label{giunzione}
\ee
\par
In the next section we consider gravitation with a minimally
coupled scalar field and study under what conditions one obtains
a collapsing shell-like solution.
Once this has been achieved, we shall employ the ADM formalism in the 
Born-Oppenheimer approach to the WDW equation as was done for 
the sphere \cite{cv,cfv} and examine the effect of back-reaction
due to particle creation on the shell during the collapse
(section~\ref{inva}).
This of course is tantamount to relaxing the adiabatic approximation.
Lastly in section~\ref{conclusion} our results are summarized and
discussed.
\section{The shell structure}
\setcounter{equation}{0}
\label{shell-like}
The purpose of this section is to illustrate how one can describe 
a collapsing matter shell in terms of a scalar field.
In particular we shall investigate the conditions under which it is
possible to neglect the thickness of the shell and use the thin limit
Hamiltonian in Eq.~(\ref{hamiltonian}).
\par
Let us begin by considering an exterior Schwarzschild metric
as given in Eq.~(\ref{metric}) with $A_{out}$.
The action for a massive scalar field $\Phi$ on such a background is
\be
S=\frac{1}{2}\,\int dt\,dr\,r^2\,
\left[\left(1-\frac{2\,G\,M}{r}\right)^{-1}\,
\left\vert \frac{d\Phi}{dt} \right\vert^2
-\left(1-\frac{2\,G\,M}{r}\right)\,
\left\vert\frac{d\Phi}{dr} \right\vert^2 
-\mu^2\,\left\vert\Phi\right\vert^2 \right]
\ ,
\label{phi_action}
\ee
where $\mu$ is the inverse Compton wave-length of the scalar
particle described by the field.
The associated Klein-Gordon equation satisfied by a spherically symmetric
(s-wave) scalar field solution $\Phi(r,t)=\varphi(r)\,e^{-i\,\omega\,t}$ 
corresponding to a collapse is
\be
\left[-\frac{d}{dr}(r^2-2\,G\,M\,r)\,\frac{d}{dr}+\mu^2\,r^2 -
\frac{\omega^2\,r^3}{r-2\,G\,M}\right]\varphi=0
\ .
\ee
On taking $r\gg R_H$, one obtains
\be
\left[-\frac{d^2}{dr^2}-\frac{2}{r}\,\frac{d}{dr}+(\mu^2-\omega^2)+
\frac{2\,G\,M}{r}\,(\mu^2-2\,\omega^2)\right]\varphi=0
\ ,
\label{hydrolike1}
\ee
which of course resembles the hydrogen atom case.
\par
The classical limit for the Coulomb case has been studied in detail
\cite{Rowe}.
In particular for large principal quantum number $n$ and angular
momentum $l=n-1$ one obtains a probability distribution strongly peaked 
on classical circular orbits.
For the case of a collapse ($n$ large and $l=0$) the probability 
distribution resembles a spherically symmetric collection of degenerate 
ellipses.
This is analogous to the harmonic oscillator where for large quantum 
numbers the position probability density similarly approaches that for 
a classical oscillator of the same energy \cite{Schiff}.
This naturally suggests, in analogy with the harmonic oscillator, that 
one may consider a coherent state, that is a wave packet that will describe
the collapse without spreading, at least for a sufficiently large period 
of time.
Unfortunately for the hydrogen atom such a construction has not yet been
successfully done (see {\em e.g.} \cite{Klauder,Bellomo}).
\par
An alternative way is to construct a shell composed of a suitable
number of (isotropic $l=0$) particles, bound together by gravitational 
interaction, whose mean position will follow the desired classical
collapse. 
In order to illustrate such an approach let us first consider the 
semi-classical (WKB) limit in Eq.~(\ref{hydrolike1}) by setting
$\varphi\sim e^{i\,S/\hbar}$.
Retaining terms to lowest order in $\hbar$ one has
\be
\left(\frac{dS}{dr}\right)^2+\hbar^2\,(\mu^2-\omega^2)+
\frac{2\,G\,M}{r}\,\hbar^2\,(\mu^2-2\,\omega^2)=0
\ .
\ee
On setting $(dS/dr)^2=p^2$ ($\ll \hbar^2\mu^2)$ one obtains
\be
\frac{p^2}{2\,(\hbar\mu)}-\frac{G\,M\,(\hbar\mu)}{r}=\hbar^2\,
\frac{(\omega^2-\mu^2)}{2\,(\hbar\mu)}
\ ,
\label{clasnewt}
\ee
which for $M=(\hbar \mu)/2$ is just the Newtonian approximation to the 
equation of motion for the collapse of a self-gravitating shell made 
of one scalar particle in s-wave with mass $\hbar\mu$ and potential
energy $G\,(\hbar\mu)^2/2\,r$ (see Eq.~(\ref{giunzione})). 
\par
The r.h.s. of Eq.~(\ref{clasnewt}) is the (non-relativistic) energy
for the motion of a particle in a gravitational field. 
We can generalize this to the case of $N$ microshells, each one made
of one s-wave particle with the same rest mass $\hbar\mu$, constituting
a self-gravitating macroshell whose mean radius satisfies a collapse 
equation similar to that for a shell of total mass $\hbar\,\mu\,N$. 
We shall see that the essential difference with respect to one shell
is that the $N$ microshells form a radially localized bound state 
corresponding to a self-gravitating macroshell of finite thickness.
\par
Let us start by considering the two microshell case.
Each microshell, with mass $\hbar\mu$ and radial position $r_i$, has
a potential energy which depends on the self-interaction plus the
interaction with the inner microshell (if present).
Hence the total potential energy can be written as
\be
V(r_1,r_2)=-\frac{G\,(\hbar\mu)^2}{2\,r_1}-
\frac{G\,(\hbar\mu)^2}{2\,r_2}-G\,(\hbar\mu)^2
\left[\frac{\theta(r_1-r_2)}{r_1}+\frac{\theta(r_2-r_1)}{r_2}\right]
\ ,
\label{pot2}
\ee
where $\theta$ is the step function.
Since we are interested in a thin macroshell we introduce the mean radius
coordinate $R=(r_1+r_2)/2$ and the relative coordinate $r=r_1-r_2$ 
with $|r|\ll R$.
In this approximation it is possible to derive for the potential energy
the following expression
\be
V(R,r)=-\frac{G\,(2\hbar\mu)^2}{2\,R}+\frac{G\,(\hbar\mu)^2}{2\,R^2}\,|r|
\ .
\ee
For realistic values of the total mass of the macroshell 
(which means that if we wish to consider physical quanta of mass 
$\hbar\mu$ we must go to a many particle description with more
particles on each microshell) the bound state has a characteristic 
time of oscillation which is several orders of 
magnitude smaller than the time needed to have a significant
variation of the centre of mass position. 
For this reason it is possible to factorize the mean radius motion 
and study the quantum bound state adiabatically with respect to the slowly 
varying $R(t)$.
An interesting fact worth noting is the (slow) increase of the coupling
during the collapse (since $R$ decreases) which implies that the average
shell thickness will decrease.
For $R$ constant the quantum problem for the relative $r$ degree of
freedom is well known and can be solved, with the spectrum
of the allowed energy excitations being given in terms of the zeros of the 
Airy function or of its derivative.
It is also possible to check that a semi-classical (WKB) analysis gives
a good estimate of the energy levels, even for low quantum numbers.
\par
A similar conclusion to the above can be obtained in a classical 
general relativistic framework by using the junction conditions 
\cite{Israel} and taking the Newtonian limit.
For a system of two non-crossing shells one has
\be
\left(\frac{dr_1}{dt}\right)^2&=&\left(1-\frac{2\,G\,M_1}{r_1}\right)
\left[-1+\frac{M_1^2}{m_1^2}+\frac{G\,M_1}{r_1}
+\frac{G^2\,m_1^2}{4\,r_1^2}\right]
\nonumber \\
\left(\frac{dr_2}{dt}\right)^2&=&\left(1-\frac{2\,G\,M_1}{r_2}\right)
\left[-1+\frac{(M_2-M_1)^2}{m_2^2}+\frac{G\,(M_1+M_2)}{r_2}+
\frac{G^2\,m_2^2}{4\,r_2^2}\right]
\ ,
\label{junc}
\ee
where $m_1,m_2$ are the proper masses of the shells positioned at 
$r_1<r_2$, $M_1$ and $t$ are the Schwarzschild mass and time coordinate
for the metric between the two shells while $M_2$ is the Schwarzschild
mass for the metric outside the outer shell.
Since we wish to analyze the interaction between the shells in the
Newtonian limit we neglect the terms proportional to $G^2$ in the above.
On taking $m_1=m_2=\hbar\mu$ one may sum (to this order) 
the two junction equations, multiply the result by $\hbar \mu/2$ and 
include the two position independent terms in the total energy $E$.
One then has, for $r_1<r_2$,
\be
E=\frac{1}{2}\hbar\mu\,\left[\left(\frac{dr_1}{dt}\right)^2
+\left(\frac{dr_2}{dt}\right)^2\right]-\frac{G\,(\hbar\mu)^2}{2\,r_1}-
\frac{3}{2}\frac{G\,(\hbar\mu)^2}{r_2}
\ .
\ee
An analogous term is easily obtained for the case $r_1>r_2$ and finally the
potential in Eq.~(\ref{pot2}) is re-obtained.
\par
If one wishes to consider a collapse approaching the horizon it is 
convenient to use the proper time of one of the two shells in 
Eq.~(\ref{junc}).
In fact it is easy to check that on the horizon the ratio of the proper
times of the two shells is $1$ up to corrections of order $d/R$ which
we require to be negligible (see Eq.~(\ref{corr}) for an estimate of this 
quantity).
This implies that one is also allowed to consider the proper time 
associated with the mean shell radius.
\par
Let us now examine the more interesting $N$ microshell case, again 
in the Newtonian limit.
One may consider each microshell at a radial position $r_i$ and define
$r_{ij}=r_i-r_j$.
The full interaction potential can then be derived:
\be
V(r_1,\dots, r_N)=-\sum_{i=1}^N\frac{G\,(\hbar\mu)^2}{2\,r_i}
-\sum_{i=1}^N\frac{G\,(\hbar\mu)^2}{2\,r_i}\,\sum_{j\ne i}
\theta(r_{ij})
\ ,
\ee
where the first sum refers to the gravitational self-interaction and
the second sum to the interparticle interaction.
Introducing the mean shell radius $R=(\sum_i r_i)/N$ one can 
write $r_i=R-(\sum_{j\ne i}r_{ji})/N$ and on using the thin shell 
approximation $|r_{ij}| \ll R$ one has
\be
V(r_1,\dots, r_N)=-\frac{G\,(N\hbar\mu)^2}{2\,R} 
+ \frac{G\,(\hbar\mu)^2}{2\,R^2}\sum_{i<j}|r_{ij}|
\ .
\ee
In this expression for the potential energy the first term indicates
that the mean shell radius motion follows approximately the Newtonian 
collapse of an object (the macroshell) of mass $m=N\,\hbar\mu$ and the 
second term refers to the gravitational interaction of the internal 
degrees of freedom of the macroshell (relative positions of the 
microshells).
One may study in the adiabatic approximation the quantum mechanical
behaviour of the internal motion of the microshells but the problem
is still complex and apparently unsolvable.
It corresponds to the case of many particles with a confining potential
linearly growing with the interparticle distances.
This of course is analogous to the quark confining potential
in hadronic physics, however for the latter case the total charge
(colour) is zero.
Further, due to the bosonic nature of the particles, in our case 
a condensate should be expected.
\par
At later stages of the collapse, as $R(t)$ varies more rapidly,
one expects non-adiabatic effects leading to transitions from the
ground to higher (excited) states for some of the constituents of the
particle condensate forming the macroshell.
Clearly such excited states may decay emitting escaping radiation.
However to describe this would mean introducing interaction terms and 
having an exterior Vaidya metric thus greatly complicating our simple
model.
The attitude we shall take and implement in the next sections is that
once enough particles in the condensate are excited (thus widening
the macroshell)
they will collectively decay to the ground state by creating 
additional particles \cite{again}.
Thus the macroshell during its collapse will be globally described as a
condensate of particles (all in the ground state) whose number varies
(increases).
\par
In order to understand how the number $N$ of microshells affects the 
global internal energy and the thickness of the macroshell
we used a many body Hartree approach to evaluate the energy for some 
trial parameter-dependent vacuum wave function.
The average Hamiltonian for a fully symmetric state
$\prod_i^N \varphi(r_i)$ is given by
\be
\expec{\hat H}&=&{\hbar^2\over2\,\hbar\mu}\,N\,\int dr\,
\left|{d\varphi(r)\over dr}\right|^2
-{G\,(N\hbar\mu)^2\over 2\,R}
\nonumber \\ 
&&+{N\,(N-1)\,G\,(\hbar\mu)^2\over 4\,R^2}\,
\int dr\,dr'\,|\varphi(r)|^2\,|r-r'|\,|\varphi(r')|^2
\ ,
\ee
where we consider 
$\varphi(r)=c\,e^{-\alpha\,|r-R|^p}\,e^{i\,k\,R}$ for the trial vacuum 
wave function with $c$ a normalization factor and 
the phase factor describes the mean shell radius behaviour which 
is approximately free.
Let us note that the asymptotic behaviour of the exact wave function for 
the $N=2$ case is given by $p=3/2$.
One can verify that minimizing the average Hamiltonian for the relative 
motion with respect to $\alpha$ leads, independently of $p$, to the 
following qualitative behaviour
\be
\expec{{\hat H}_{rel}}\sim \hbar\mu\,N\,(N-1)^{2/3}\,
\left(\frac{G}{R^2}\right)^{2/3}
\ ,
\label{hartres}
\ee
and $p$ only affects the magnitude of the proportionality constant.
Further one also obtains an estimate of the width $d$ of the packet
\be
d\sim\alpha^{-1/p}
\sim\frac{1}{\mu}\,\left(\frac{G\,(N-1)}{R^2}\right)^{-1/3}
\ .
\label{width}
\ee
\par
A similar behaviour can also be found if one calculates the energy
levels semi-classically. 
Indeed for such a case the potential term may be evaluated by considering 
the microshells to be well ordered \cite{Nambu} so that the positions 
$\tilde r_k =r_k-R$ (with respect to the mean shell radius) satisfy
$\tilde r_1 \le \tilde r_2 \le \dots \le \tilde r_N$ 
and one will have $r_{k+1,k}\ge 0$ with
\be
\sum _{k=1}^N \sum_{j<k} |r_{kj}| = 
\sum_{k=1}^{N-1}\,k\,(N-k)\,r_{k+1,k}=
2\,\sum_{k=1}^N\,k\,\tilde r_k
\ .
\ee
The $N$ microshell Hamiltonian then becomes
\be
H=\sum _{k=1}^N H_k= \sum_{k=1}^N \left[\frac{1}{2}\,\hbar\mu\, 
\dot {\tilde r}_k^2
+\frac{G\,(\hbar \mu)^2}{2\,R^2}(2\,k) \tilde r_k\right]
\ .
\ee
We may now consider the $k$-th microshell whose amplitude 
of oscillation will be the width $d$ of the shell. 
Taking a closed cycle for the $k$-th microshell one has
a semiclassical quantization condition:
\be
4\,\hbar\mu\,\sqrt{\frac{G\,\hbar\mu}{R^2}\,(2\,k)} 
\int_0^d (d-\tilde r _k)^{1/2}\, d\tilde r_k=
\left(n+\frac{1}{2}\right)\,\hbar
\ ,
\label{quantcond}
\ee
which leads to an energy
\be
E_k=\left(\frac{9}{128}\,\hbar\mu\right)^{1/3}\,
\left[\frac{G\,\hbar\mu}{2\,R^2}(2\,k)\,
\left(n+\frac{1}{2}\right)\,\hbar\right]^{2/3}
\ee
and
\be
E=\sum_{k=1}^N E_k \approx 
\frac{N^{5/3}}{16}\,\hbar\mu\,\left[\frac{3}{2}
\,\frac{G}{R^2}\,\left(n+\frac{1}{2}\right)\,\hbar\right]^{2/3}
\ ,
\ee
in agreement with Eq.~(\ref{hartres}). 
We note that the binding energy increases faster that the microshell
number.
In particular for a given $n$ the smaller is $R$ and/or the larger is 
the number $N$ of microshells the smaller is the thickness $d$ of the 
macroshell.
\par
Using natural units $\hbar=c=1$ one has $G=1/m_{pl}^2\simeq 10^{-38}$
GeV$^{-2}$ and we can obtain some numerical estimates.
If we consider matter quanta with a mass $\hbar\mu=1$ GeV and a number of 
quanta $N=10^{51}$ (giving a total mass corresponding to 
$10^{-6}$ solar masses), the Schwarzschild radius will be 
$2Gm= 10^{13}\,\mu^{-1}\simeq 0.2$ cm and for a mean shell radius 
located at $R=10^8\,G\,m\simeq 200$ km, one has, from Eq.~(\ref{width}), 
an estimated thickness of the macroshell 
$d\sim 10^{10}\,\mu^{-1}\sim 10^{-4}$ cm.
With these numbers the potential energy of the mean radius of the 
macroshell is $G\,m^2/(2\,R)\sim 10^{43}$ GeV, while the total 
internal energy for the other degrees of freedom, assumed to be in the 
ground state, is $E_{int}\sim m\,(G\,N/R^2)^{2/3} \sim 10^{32}$ GeV 
(higher occupation numbers would lead to the previously shown $n^{2/3}$
behaviour).
It is clear from these numbers that the condensate-maintaining mechanism
we previously envisaged is plausible:
a limited number of excited states can easily supply the energy for
the creation of additional particles thus maintaining the condensate.
\par
From the above picture it is clear that, for a wide regime during the 
collapse, modelling the macroshell in terms of scalar fields 
confined around the mean shell radius is a good approximation.
Thus we shall take for the scalar field describing one of the $N$ 
particles a form
$\Phi(r,t)=e^{-i\,\mu\,t}\,\phi(t)\,\varphi(r)/\sqrt{\mu}$ with 
$\varphi(r)$ describing a packet with a width of order $d$ around $R$.
This time $ \varphi(r) $ will be normalized with the measure $dr$ and 
not $r^2\,dr$ since we shall consider the dynamics of the field 
$\phi(t)$ with $R(t)$ a time varying parameter.
Then, if one considers a radial wave packet similar to the one
used in the Hartree approximation and performs the radial integration
in the action (\ref{phi_action}) for the scalar field, one obtains 
an effective action for $\phi(t)$.
In terms of the proper time $\tau$ such effective action is given by
\be
S\approx\frac{1}{2\,\mu}\,\int d\tau\,R^2(\tau)\,\left[
\left(\frac{d\phi}{d\tau}\right)^2 -\mu^2\,\phi^2 
-\mu^2\,\left(\frac{G\,N}{R^2}\right)^{2/3}\,\phi^2 \right]
\ .
\label{matteraction}
\ee
The last term in the integrand, which is related to the shell thickness, 
on comparing with the second term, can be safely neglected.
Indeed, with the numbers previously given it is suppressed by a factor 
of $10^{19}$.
In the next section we shall use the formalism of second quantization
in order to describe the $N$ particle system discussed above.
\par
Lastly we note that, since we shall be interested in the evolution of 
the shell until it reaches the Schwarzschild radius $R_H$, which can be 
approximated extremely well by $2\,G\,m$, we must check that the 
ratio 
\be
\frac{d}{R}\simeq\frac{E_{int}}{m}  
\sim \left({G\,N\over{R^2}} \right)^{2/3}
\label{corr}
\ee
is still negligible with respect to 1 near $R_H$, that is 
\be
\label{c1}
   G \, N \, \mu \sim R_H \gg 1 / \mu \ . 
\ee
For the values given above ($\mu=1$ GeV), we need $N\gg 1/G\,\mu^2\sim 10^{38}$.
\section{The quantized shell}
\setcounter{equation}{0}
\label{inva}
We may now write the WDW equation for the collapsing shell and the 
scalar field.
The corresponding Hamiltonian will be given by Eq.~(\ref{hamiltonian})
with $m$ the average value for $N$ particles of the Hamiltonian for the
quantum scalar field obtained from Eq.~(\ref{matteraction}).
Let us now illustrate this.
On expanding the Hamiltonian in Eq.~(\ref{hamiltonian}) in $G$ 
and retaining terms up to order $G$ one has
\be
H={\mu\over2}\,\left[{\pi _\phi ^2\over R^2}+R^2\,\phi^2\right]
-M+{P_R^2 \over 2\,M}-{G\,M^2\over 2\,R}\equiv H_M+H_G
\ ,
\ee
where $\pi_\phi=R^2\,\dot\phi/\mu$ and $P_R=M\,\dot R/\sqrt{1+\dot R^2}$.
Canonical quantization then leads to the WDW equation
\be
\left[{\mu\over 2}\,\left(-{\hbar^2\over R^2}\, 
{\partial^2\over\partial \phi^2}+R^2\,\phi^2\right)
-M-{\hbar^2\over 2\,M}\,{\partial^2\over \partial R^2}
-{G\,M^2\over 2\,R} \right]\Psi (R,\phi)=0
\ ,
\ee                                                    
and there is no operator ordering ambiguity.
\subsection{Born-Oppenheimer approach}
We express the function $\Psi$ in the form
$\Psi(R,\phi)=\psi(R)\,\chi(\phi,R)$
which allows us to obtain the WDW equation for the gravitational 
part \cite{bertoni}
\be
\left[-\left(M+{G\,M^2\over2\,R}+{\hbar^2\over2\,M}\,
{\partial^2\over\partial R^2}\right)
+{1\over2\,\pro{\tilde \chi}{\tilde \chi}}\,\bra {\tilde \chi}\,
\left(\mu\,{\hat \pi^2_\phi\over R^2}+\mu\,R^2\,\phi^2\right)
\ket{\tilde \chi}\right]\,\tilde \psi
\nonumber \\
\equiv \left[\hat H_G + \expec{H_M} \right] \tilde \psi
={\hbar^2\over2\,M}\,{1\over \pro{\tilde \chi}{\tilde \chi}}
\bra{\tilde \chi}\,{\partial ^2\over\partial R^2}\ket{\tilde \chi}
\equiv
{\hbar ^2\over2\,M}\,\expecl{\partial^2\over\partial R^2}\,
\tilde \psi
\ ,
\label{gravitation}
\ee
where we have defined a scalar product
\be
\pro{\chi}{\chi} \equiv \int d\phi\,\chi^*(\phi,R)\,\chi(\phi,R)
\ee
and set
\be
\psi= e^{-i\,\int^R A(R^{\prime})\,dR^{\prime}}\,\tilde \psi
\ \ \ \ \ \,
\chi= e^{+i\,\int^R A(R^{\prime})\,dR^{\prime}}\,\tilde \chi 
\ee
with
\be
A \equiv -{i\over\pro{\chi}{\chi}}\,\bra{\chi} 
{\partial \over\partial R} \ket{\chi}
\equiv -i \, \expecl{ \partial \over\partial R}
\ .
\label{geometric}
\ee
Further, we also obtain the equation for the matter field
\be
\tilde\psi\,\left[\hat H _M-\expec{\hat H_M} \right]\tilde \chi
+{\hbar ^2\over2\,M}\,\left({\partial \tilde \psi\over\partial R}
\right)\,\left({\partial\tilde \chi\over\partial R}\right)
={\hbar^2\over2\,M}\,\tilde \psi\,
\left[\expecl{\partial^2\over\partial R^2}
-{\partial^2\over\partial R^2}\right]\tilde\chi 
\ .
\label{matter}
\ee
\par
As yet there is no time parameter, however, on neglecting the r.h.s. 
of Eq.~(\ref{gravitation}), one may introduce a time variable by
taking the semi-classical (WKB) approximation 
for the gravitational wave function $\tilde \psi$,
\be
\tilde\psi\simeq \left({\partial S_{eff}\over\partial R} \right)^{-1/2}\, 
e^{{i\over\hbar}\,S_{eff}}
\ ,
\ee
where the effective action $S_{eff}$ satisfies the Hamilton-Jacobi 
equation 
\be
\expec{\hat H _M} + H_G =0
\label{hj}
\ee
associated with the l.h.s. of Eq.~(\ref{gravitation}),
and in the non-relativistic limit, $\dot R\ll 1$, it is given by
\be
S_{eff}={1\over 2}\,\int d\tau\,\left[2M +M\,\dot R^2
+{G\,M^2\over R}-2\,\expec{\hat H_M} \right]
\ ,
\label{S_eff}
\ee
where $\expec{\hat H_M}$ is evaluated on the semi-classical 
trajectory where $\tilde \psi$ has support.
\par
The explicit expression of Eq.~(\ref{hj}),
\be
\dot R^2 ={2\over M}\,\left(M-\expec{\hat H_M}\right)
+{G\,M\over R}
\ ,
\ee
may be compared with the classical Eq.~(\ref{giunzione}).
The two equations lead to the same evolution if
\be
\expec{\hat H_M}=m
\ ,
\ee
and
\be
&&{G^2\,m^2\over4\,R^2} \ll {G\,M\over R}\ \ \ \Rightarrow
\ \ \  {G\,m^2\over R\,M}\ll 1
\nonumber \\
&&{M^2\over m^2}-1\simeq{2\over M}\,(M-m)
\ \ \ \Rightarrow \ \ \ m/M\simeq 1
\ .
\label{ratios}
\ee
With the above conditions we can obtain as our equation of motion
for $R$
\be
\dot{R}^2-{G\,M \over R} = {M^2 \over m^2} -1
\ ,
\label{classeq}
\ee
which is just the equation for radial geodesics in Schwarzschild 
space-time.
In particular, on comparing Eq.~(\ref{classeq}) with the standard
notation for a radial trajectory with geodesic energy $E$ in a 
Schwarzschild space-time with mass parameter $\tilde M$  
(see {\em e.g.} \cite{stephani}),
\be 
\dot R^2={2\,G\tilde M \over R}-(1-E^2)
\ ,
\ee
we find $\tilde M=M/2$ and $E=M/m$.
Hence the shell evolves as a particle in a Schwarzschild manifold 
of total energy $M/2$, that is half of the total energy of the shell.
The solutions to Eq.~(\ref{classeq}) are known. 
Here we only recall for later reference the case $m=M$, 
corresponding to the separatrix between bound 
orbits and scattering trajectories:
\be
R_c(\tau)=R_0\,
\left[1-{3\over 2}\,\sqrt{G\,M\over R_0^3}\,\tau\right]^{2/3}
\ ,
\label{R_c}
\ee
where $R_0\equiv R(0)$.
\par
The proper time variable associated to the WKB gravitational wave
function is given by the relation
\be
{\partial \tilde \psi\over \partial R}\,{\partial\over\partial R} 
&\simeq&
\left[{i\over \hbar}\,{\partial S_{eff}\over\partial R}
-{1\over 2}\,\left({\partial S_{eff}\over\partial R}\right)^{-1}\,
{\partial^2 S_{eff}\over\partial R^2}\right]\tilde \psi\,
{\partial \over \partial R}
\nonumber \\
&\equiv&-{i \over \hbar}\,M\,\tilde \psi\, 
{\partial\over\partial \tau}-{1\over 2}\,
\left({\partial S_{eff}\over\partial R}\right)^{-1}\, 
{\partial^2 S_{eff}\over\partial R^2}\,\tilde \psi\, 
{\partial\over\partial R} 
\ .
\label{wkb}
\ee
If the r.h.s. of Eq.~(\ref{matter}) and the second term in 
Eq.~(\ref{wkb}) are negligible, one obtains a Schr\"odinger equation 
for the matter wave-function $ \chi _s $,
\be
i\,\hbar\,{\partial \chi _s\over \partial \tau}
={\mu\over 2}\,\left[-{\hbar^2\over R^2}\, 
{\partial^2\over\partial \phi^2}+ R^2\,\phi^2\right]\chi_s
=\hat H_M\, \chi _s
\ ,
\label{Schrod}
\ee
where we have scaled the dynamical phase
\be
\chi_{s} \equiv\tilde\chi\,\exp\left\{-{i\over\hbar}\,\int ^{\tau} 
\expec{\hat H _M(\tau')}\,d\tau'\right\}
\ .
\label{dynamic}
\ee
\subsection{Adiabatic invariants}
In order to study quantum systems with explicitly time-dependent
Hamiltonians we can use the method of invariants \cite{lewis-ries}.
Given a Hamiltonian $ \hat H_M (\tau) $, a Hermitian operator
$\hat I(\tau)$ is an invariant if it satisfies
\be
i\,\hbar\,{\partial \hat I(\tau)\over\partial \tau} +
\left[\hat I(\tau),\hat H_M(\tau)\right] =0 
\ .
\ee
The general solution  $\chi_s(\tau)$ to the Schr\"odinger 
equation
\be
i\,\hbar\,{\partial \chi_s(\tau)\over\partial \tau} =
\hat H_M(\tau)\,\chi_s(\tau)
\ee
can then be written in the form
\be
\ket{ \chi,\,\tau}_{Is} =
\sum_n C_n\,e^{i\,\varphi_n(\tau)}\,\ket{n,\tau}_I
\ ,
\ee
where $\ket{ n, \tau }_I  $ is an eigenvector of $ \hat I (\tau) $ 
with time-independent eigenvalue $\lambda_n$ and $C_n$ are complex
coefficients.
The phase $ \varphi_n ( \tau ) $ is given by the sum of the geometrical 
and dynamical phases associated respectively to Eqs.~(\ref{geometric}) 
and (\ref{dynamic}),
\be
\varphi_n(\tau)= {i\over \hbar}\,\int_{\tau_0}^{\tau}
\ _I\,\bra{n,\tau'}\,\hbar\,\partial_{\tau'} +
i\,\hat H_M(\tau')\,\ket{n,\tau'}_I\,d\tau'
\ .
\ee
\par
The Hamiltonian in Eq.~(\ref{Schrod}) describes an harmonic oscillator
of fixed frequency $\mu$ and variable mass $ R^2 / \mu $. 
Thus we can introduce the linear (non-Hermitian) invariant 
\be
\hat I_b(\tau) \equiv e^{i\,\Theta(\tau)}\,\b(\tau)
\ ,
\ee
with the phase $ \Theta(\tau) $ given by
\be
\Theta(\tau)= \int^{\tau}_{\tau_0} d\tau'\,
   {\mu\over R^2\,(\tau')\,x^2(\tau')}
\label{theta}
\ee
and
\be
\b(\tau) \equiv {1\over{\sqrt{2\,\hbar}}}\,\left[
{\hat \phi\over x}+i\,\left(x\,\hat\pi_{\phi}-\dot x \, 
{R^2\over\mu}\,\hat\phi\right) \right] 
\ .
\ee
The function $ x(\tau)$ is a solution of
\be
\ddot{x}+2\,\dot x\,{\dot R \over R} +\mu^2\, x
={\mu^2\over R^4\,x^3}
\ ,
\label{x}
\ee
with suitable initial conditions.
\par
The system admits an invariant ground state (vacuum) $\ket{0,\tau}_b$ 
defined by
\be
\hat I_b(\tau)\,\ket{0,\tau}_b =0
\ ,
\ee
from which one can define a basis of eigenstates 
${\cal B}=\{\ket{n,\tau}_b\}$,
\be
\ket{n,\tau}_{bs} &\equiv& 
{(\hat I^{\dagger}_b)^n \over\sqrt{n!}}\,\ket{0,\tau }_{bs}
\nonumber  \\
&=&e^{-i\,n\,\Theta}\,{(\hat b^{\dagger})^n\over \sqrt{n!}}\,
\ket{0,\tau}_{bs}
\nonumber \\
&=&e^{i\,(\varphi_0-n\,\Theta)}\,{(\hat b^{\dagger})^n\over \sqrt{n!}}
\,\ket{0,\tau}_b=e^{i\,\varphi_n(\tau)}\,\ket{n,\tau}_b
\ ,
\label{n-states}
\ee
where $\varphi_0 $ is arbitrary and can be replaced by
$\Theta/2$, and
\be
&\b\,\ket{ n, \tau }_b = \sqrt{n}\,\ket{ n-1,\tau}_b&
\nonumber \\
&\bc\,\ket{ n, \tau }_b = \sqrt{n+1}\,\ket{ n+1,\tau}_b
\ .&
\ee   
Since $\left[\b,\bc\right]=1$ we will refer to  $\bc$ and $\b$ as
the invariant creation and annihilation operators.
We can then introduce the Hermitian quadratic invariant operator
\be
\hat I_c \equiv \left(\hat I_b^{\dagger}\,\hat I_b + 
{1 \over 2} \right) =\hbar\,\left(\bc\,\b + {1 \over 2}\right)
\ ,
\ee
in which $\bc\,\b$ is the invariant number operator.
\par
We also define the particle creation and annihilation operators
$\ac$ and $\a$ (with $\left[ \ac,\a \right]=1$) as 
\be
\a(\tau) &=&
{R\over \sqrt{2\,\hbar}}\,\left(
\hat \phi +i\,{\hat \pi_{\phi}\over R^2} \right) 
\nonumber  \\
\ac(\tau) &=&
{ R\over \sqrt{2\,\hbar}}\,\left(
\hat \phi -i\,{\hat \pi_{\phi}\over R^2} \right) 
\ ,
\ee
in terms of which the matter Hamiltonian operator can be expressed as
\be
\hat H_M = \hbar\mu\,\left(\hat N + {1 \over 2} \right)
\ ,
\ee
where $ \hat N \equiv \ac\,\a $ is the particle number operator
which counts the number of quanta of the scalar field $\phi$.
The corresponding vacuum state $ \ket{0,\tau}_a $ is defined by
\be
 \a\,\ket{0,\tau}_a=0 
\ee
and a complete set of eigenstates can be given as
${\cal A} = \left\{ \ket{n,\tau}_a \right\} $,
\be
\ket{n,\tau}_a \equiv
{ {\left( \ac \right) ^n}\over{\sqrt{n!}}} \, \ket{0,\tau}_a
\ ,
\ee
with
\be
&\a\,\ket{n,\tau}_a = \sqrt{n} \, \ket {n-1,\tau}_a & 
\nonumber \\
&\ac\,\ket{n,\tau}_a = \sqrt{n+1} \, \ket{n+1,\tau}_a 
\ . &
\ee
\par
The two Fock basis  $ {\cal A} $ and $ {\cal B} $ are related.
On using
\be
&\hat\phi = \sqrt{\hbar\over 2 }\,x\,\left(\b +\bc\right) &
\nonumber \\
&\hat \pi_{\phi}= \sqrt{\hbar\over 2}\,\left[{i\over x}\, 
\left(\bc -\b\right)+\dot x\,{R^2\over \mu}\,\left(\bc+\b\right)
\right] 
\ ,&
\ee
we obtain
\be
\begin{array}{lcr}
\left\{
\begin{array}{l}
   \a =B^{*}\, \b + A^{*}\, \bc   
\\
   \ac = B\, \bc +A\, \b
\end{array} \right.
&{\rm or}&
\left\{
\begin{array}{l}
   \b = B\, \a - A^{*}\, \ac
\\
   \bc= B^{*}\, \ac -A\, \a
\ ,
\end{array}
\right.
\end{array}
\label{Bogoliubov}
\ee
where
\be
   A(\tau) = {1 \over 2} \, R\, \left(
   x - { 1 \over { R^2 \, x}} -i \, { {\dot x}\over \mu}
   \right)  
\nonumber \\
   B(\tau)= { 1 \over 2} \, R\, \left(
   x + { 1 \over {R^2 \, x}} -i \, {{\dot x} \over \mu}
   \right)
\ee
are the Bogoliubov coefficients.
The two basis $ {\cal A} $ and $ {\cal B} $ will coincide
for $ \tau = \tau _0 $ if
\be
\b(\tau _0 )= \a (\tau_0)\ \ \ \Rightarrow\ \ \ 
\mu\, \hat I _c( \tau _0 ) = \hat H_M (\tau_0)
\ ,
\ee
that is $ A (\tau_0)=0$ and $ B(\tau_0)=1$,
which means that for the function $ x(\tau) $ we must require
\be
\left\{\begin{array}{l}
x(\tau_0)= 1/R 
\\
\dot x (\tau_0)= 0 
\ .
\end{array}\right.
\ee
In the following we shall take $ \tau_0 = 0 $ so that the adiabatic 
approximation holds at the beginning of the collapse.
\par
If we define $\sigma \equiv R \, x$, Eq.~(\ref{x}) takes the form
\be
{{\ddot{\sigma}} \over \mu^2} + \Omega^2\, \sigma=
  {1 \over \sigma^3}   
\ ,
\label{pinney}
\ee
which is the equation for a harmonic oscillator
with time-dependent frequency $ \Omega $,
\be   
\Omega^2 =1+{{\ddot R}\over{\mu^2 R^2}} 
\ .
\ee
On using the equation of motion (\ref{classeq}) one has
\be
  \Omega ^2 \simeq 1+{G\,M \over 2\,\mu^2\,R^3} \equiv 
  1 + {\delta ^2 \over 2} \, {R_0^3  \over R^3} \ ,
\ee
where $ \delta $ (also suggested by Eq.~(\ref{R_c})) plays the role of 
a parameter measuring non-adiabaticity since for $ \delta \to 0 $ the 
adiabatic limit is recovered.  
On expanding in $ \delta $, an approximate analytic solution to 
Eq.~(\ref{pinney}) is given by \cite{lewis}
\be
\sigma \simeq \Omega^ {-1/2}=1-{\delta ^2 \over 8} \, {R_0^3  \over R^3}
\ .
\ee
Then
\be
x={ \sigma \over R} \simeq{1 \over R} \,
    \left[1- {\delta ^2 \over 8} \, {R_0^3  \over R^3} \right]
\label{xx}
\ee
and
\be
   \Theta (\tau) \simeq \mu\, \int_0^{\tau} 
	d\tau' \, 
       \left[1+ {\delta ^2 \over 4} \, {R_0^3  \over R^3} \right]
\ .
\ee
\par 
The introduction of the eigenstates of adiabatic invariants 
allows us to examine the particle production induced by 
the evolution of the radius of the shell.
As we have seen previously, in order to have a physical shell
we must consider a large number of particles.
Thus let us consider a linear combination of eigenstates of the 
invariant number operator defined in Eq. (\ref{n-states}),
\be
   \ket{N_\lambda,\tau} _{bs} = {1 \over \sqrt{2\lambda}} \,
   \sum _{n=N-\lambda}^{N+\lambda}
   \ket {n,\tau}_{bs}
   \ ,
\ee
with $ 1 \ll \lambda \ll N $.
Then we can evaluate the quantities of interest as 
expectation values for this state.   
On defining $\expec{\hat O}\equiv\
_{bs}\bra {N_\lambda,\tau}\hat O\ket{N_\lambda ,\tau}_{bs}$
we obtain (up to corrections of ${\cal O}(\lambda/N)$)
\be
   \expec {\phi} &\simeq& \sqrt{2\, \hbar \, N}\, x\, \cos \Theta
 \nonumber \\
   \expec { \pi_{\phi} } &\simeq&  \sqrt{2\, \hbar\, N}
   \left[ {\dot x\, R^2 \over \mu} \cos \Theta - { 1 \over x}\,\sin \Theta
   \right]
\ee
and
\be
\expec{\left(\Delta\hat\phi\right)^2}
&\simeq&  { \hbar \over 2 } \, x^2 
\nonumber  \\
\expec{\left(\Delta\hat\pi_{\phi}\right)^2}
&\simeq&  {\hbar\over 2}\,\left({1\over x^2} + {R^4 \over \mu ^2} 
  \dot x^2 \right) 
\ ,
\ee
which lead to an uncertainty relation
\be
\expec{\left(\Delta \hat \phi \right)^2}^{1/2}\,
\expec{\left(\Delta \hat \pi_\phi\right)^2}^{1/2}
\simeq{\hbar \over 2}\, \sqrt{1+{R^4 \over \mu ^2}\,x^2\,\dot x^2}
\ .
\ee
On substituting the explicit expression for $ x $ and $ \dot x $
from Eq.~(\ref{xx}) we obtain 
\be
\expec{\left(\Delta \hat \phi\right) ^2 }&\simeq&
{\hbar \over 2}\,{1 \over R^2}\,
\left(1- {\delta^2 \over 4} \, {R_0^3  \over R^3} \right)
\nonumber    \\
\expec{\left( \Delta \hat \pi_\phi\right) ^2}&\simeq&
{\hbar \over 2}\,\left[R^2\,
\left(1+ {\delta ^2 \over 4} \, {R_0^3  \over R^3}\right) 
+{\dot R ^2 \over \mu^2}\,
\left(1-\delta^2  \, {R_0^3  \over R^3}\right)\right]
\ ,
\ee
and
\be
\expec{\left(\Delta \hat \phi \right)^2}^{1/2}\,
\expec{\left(\Delta \hat \pi_\phi\right)^2}^{1/2}
\simeq{\hbar\over 2}\,\left[1+{\dot R^2\over\mu^2\,R^2}\,
\left(1-{5\over 4}\,\delta^2\,{R_0^3\over R^3}\right)\right]^{1/2}
\ .
\ee
We also find
\be
   \expec { \hat I _c} \simeq \hbar\,\left( N+ { 1 \over 2} \right)
\ee
and the expectation value of the Hamiltonian is
\be
\expec{\hat H _M} \simeq
{\mu \over 2R^2} \left[ {\hbar \over 2} \left( x^2 R^4 + {1 \over x^2}
+ \dot x ^2 {R^4 \over \mu ^2} \right) + \expec {\hat \pi _{\phi}} ^2
+R^4 \expec { \hat \phi} ^2 \right] \ .
\ee
\par
On approximating 
\be
\ket{N_\lambda,\tau}_{bs} \sim \ket{ N,\tau}_{bs}
\label{N_l}
\ee
we have
\be
\expec{\hat H _M} &\simeq&
{\mu \hbar\over 2}\,\left(N+{1 \over 2}\right)\,
\left[{1\over x^2\,R^2}+x^2\,R^2+{\dot x^2\,R^2 \over\mu^2}\right] 
\nonumber \\
&\simeq& \hbar\mu\,N\,
\left[1+{\dot R^2\over 2\,\mu^2\,R^2}\right]
\nonumber \\
&=& m_0\,\left[1+ {\dot R^2\, R_0^3 \over R_H\, R^2}\,\delta^2 \right]
\nonumber \\
&\equiv& m_0+\Delta m(\tau)
\ ,
\label{mass}
\ee
to leading order in $ \delta $.
This value can be identified with a varying proper mass $ m(\tau) $
of the shell.
\subsection{Consistency conditions and back-reaction}
\label{simulation}
The previous results were obtained on neglecting the r.h.s.s and the
higher WKB order terms in the equations for the matter and gravitational 
wave-functions.
Thus, in order to check the validity of our results, 
we must estimate the neglected terms and compare them to the
corrections we obtain with respect to the classical trajectory
in Eq.~(\ref{R_c}).
In the following it will be sufficient to consider states 
$\ket{N_\lambda,\tau}_{bs}$ in the approximation (\ref{N_l}) and study 
their evolution for the mean shell radius $R$ greater than the 
Schwarzschild radius $R_H$. 
\par
The r.h.s. of the gravitational equation (\ref{matter}) can be
estimated according to
\be
{\hbar^2\over 2\,M}\,
_b\bra{N}\partial^2_R\ket{N}_b &=&-{\hbar^4\mu^2\over 4\,M\,R^2}\,
\sum_{P\neq N}{\bra{N}({\hat a}^{\dagger 2}+{\a}^2)\ket{P}\,\bra{P}\,
({\hat a}^{\dagger 2}+{\a} ^2)\ket{N} \over (E_N-E_P)^2 }     
\nonumber \\
&\simeq& -{m^2\over 4\,\mu^2\,M\,R^2}  
\ ,
\label{rhsg}
\ee
where we used for the time-dependent number of scalar quanta
the relation $N(\tau)=m/\hbar\mu$. 
Therefore, the above correction to the Hamilton-Jacobi equation for 
gravity is negligible with respect to the matter source 
$m=\expec{\hat H_M}$ if
\be 
R\gg {1/\mu} 
\ ,
\label{condition1}
\ee
which means that the fluctuations associated with the quantum nature of
matter become dominant for a radius of the shell of the order of the 
Compton wave-length of the constituent quanta 
(which appears as a general property of quantized models, see also 
Refs.~\cite{cv,cfv}).
It is worth noting that the condition (\ref{c1}) which ensures 
that the width of the shell $ d \ll R $ coincides with
(\ref{condition1}) when $ R=R_H $.
\par
For suitable initial conditions it is possible to have $R>R_H\gg 1/\mu$,
so that one may follow the evolution of the shell down to its
Schwarzschild radius while using the semi-classical Hamilton-Jacobi equation.
However, on using the second line in Eq.~(\ref{mass}) to estimate
$\Delta m$ one finds that the ratio $\rho$ between $\Delta m$
and the quantity in Eq.~(\ref{rhsg}) is
\be
\rho<2M\,{m_0\over m^2}
\ ,
\ee
where the maximum (of order 2) is obtained on the horizon.
Hence, in order to compute consistently the back-reaction on the 
trajectory $R(\tau)$ due to an increasing number of quanta, 
one must include the contribution in Eq.~(\ref{rhsg}) into 
the Hamilton-Jacobi equation.
This leads to the equation of motion given below by the first 
expression in Eq.~(\ref{R_s}).
\par
The r.h.s. of the matter equation (\ref{matter}) induces quantum
transitions from the (otherwise conserved) state $\ket{N}_b$
to states $\ket{L}_b$ with $L\neq N$.
It is then useful to estimate the projection 
\be
_b\bra{L}\partial^2_R\ket{N}_b &=&  
{\hbar^2\mu^2\over R^2}\,\sum_{P\neq L,N}
{\bra{L}({\hat a}^{\dagger 2}+\a^2)\ket{P}\,
\bra{P}({\hat a}^{\dagger 2}+\a ^2)\ket{N} 
\over(E_P - E_L)\,(E_P-E_N)}
\nonumber \\
&\simeq&{m^2\over 4\,\mu^2\,R^2}
\ ,
\ee
for $L=N\pm 4$.
Since the latter expression leads to a r.h.s. in Eq.~(\ref{matter})
of the same order of magnitude as Eq.~(\ref{rhsg}), it also leads to 
the same condition, Eq.~(\ref{condition1}), for the validity of the 
Schr\"odinger equation (\ref{Schrod}).
Further, in Eq.~(\ref{Schrod}), if one wishes to include corrections 
related to $\Delta m$, one must also include the contribution of the 
r.h.s. of Eq.~(\ref{matter}) as was done above for the gravitational
part.
In the following we shall check that, for physically plausible data,
$\Delta m\ll m_0$ and both the above corrections can be neglected. 
\par
Finally we must check the consistency of the WKB approximation used
in Eq.~(\ref{wkb}).
For the matter equation (\ref{matter}) the correction coming from 
the (neglected) prefactor in the semi-classical wave function 
$\tilde\psi$ is approximatively given by
\be
-{\hbar^2\over2\,M}\,{\dot P_R\over\dot R\,P_R}
\sim {\hbar^2\,\ddot R\over 2\,M\,\dot R^2}
\le{\hbar^2\over G\,M^2}
\ ,
\ee
where the last expression is the value on the horizon $R_H$.
For the gravitational equation the corresponding correction is 
given by
\be
{\hbar^2\over2\,M}\,\left[-{3\over 4}\,{\dot P_R^2\over\dot R^2\,P_R^2}
-{\dot P_R\,\ddot R\over2\,\dot R^3\,P_R} 
+{\ddot P_R\over2\,\dot R^2\,P_R}\right]
\sim 
{\hbar^2\over2\,M}\,\left[
-{5\,\ddot R^2\over 4\,\dot R^4}+\frac{1}{2 \,\dot R^3}\mathop{R}^{\dots}
\right] \le {\hbar^2\over G^2\,M^3}
\ .
\ee
Both corrections above, when compared with the matter source $m$ are
negligible provided
\be
R_H\gg{\hbar/ M}
\ ,
\label{condition2}
\ee
where $\hbar/M$ is the Compton wave-length of the shell as
a whole and is much smaller than $1/\mu$ (see Ref.~\cite{cv}).
Thus if condition (\ref{condition1}) is satisfied, so is 
(\ref{condition2}).
Further, the first correction is proportional to $N^{-2}$ and 
the second one to $N^{-3}$. 
From the expression for $m$ in Eq.~(\ref{mass}) one can infer 
that $\Delta m$ is proportional to the ratio between the Compton 
wave-length $1/\mu$ and the radius at which $\Delta m$ is computed 
(in our case $R>R_H$).
Thus $\Delta m\sim 1/N$ (on the horizon where $\dot R$ is a constant) 
and both corrections are negligible with respect to $\Delta m$ 
for $N\gg 1$.
\par
In order to compute the back-reaction on the trajectory of the shell,
one can integrate the system of coupled ordinary differential equations 
governing the semi-classical (Newtonian) trajectory, which we denote 
by $R_s$, and the corresponding proper mass $m$,
\be
\left\{\begin{array}{l}
\dot R_s^2={G\,M\over R_S}+2\,\left(1-{m\over M}\right)
-{m^2\over 4\,\mu^2\,M^2\,R_s^2}
\\
m=m_0\,\left(1+{\dot R_s^2\over 2\,\mu^2\,R_s^2}\right)
\ .
\end{array}\right.
\label{R_s}
\ee
The last term in the first equation above corresponds to the quantity
in Eq.~(\ref{rhsg}).
As initial condition we shall set $m(0)\equiv m_0=M$. 
It is then natural to compare the result with $R_c$ in Eq.~(\ref{R_c})
which corresponds to the Newtonian limit for the trajectory of a thin 
shell with $m=m_0=M$ constant.
\par
We have computed $R_s(\tau)$ and $m(\tau)$ numerically for a set
of initial number of quanta $N > 10^{38}$, which is the limit required
by our model of the shell described in section~\ref{shell-like}
(see Table~\ref{table1}, the third case is included only as
a limiting example).
\begin{table}
\centerline{\begin{tabular}{|c|c|c|c|c|c|}
\hline
N & $R_H$ & $\Delta N$ & $R_s-R_c$ & ${R_s-R_c\over R_c}$ &
${m^2\over 8\,M^2}$ \\
\hline
$4\cdot10^{40}$ & $500\,{1\over\mu}$ 
& $4\cdot10^{34}$ & $2\cdot10^{-3}\,{1\over\mu}$ 
& $4\cdot 10^{-6}$ & 0.13 \\
\hline
$4\cdot10^{39}$ & $50\,{1\over\mu}$
& $4\cdot10^{35}$ & $2\cdot10^{-2}\,{1\over\mu}$ 
& $4\cdot 10^{-4}$ & 0.13 \\
\hline
$4\cdot10^{38}$ & $5\,{1\over\mu}$
& $4\cdot10^{36}$ & $2\cdot10^{-1}\,{1\over\mu}$ 
& $4\cdot 10^{-2}$ & 0.13
\\
\hline
\end{tabular}}
\caption{Samples of numerical estimates of the effects associated 
with non-adiabaticity.
$N$ is the initial number of particles (with Compton wave-length 
$1/\mu=2\cdot10^{-14}$~cm) in the shell; $R_H$ the corresponding 
Schwarzschild radius; $\Delta N$ the number of particles produced 
at the proper time at which the shell coordinate $R_c$ reaches 
the horizon; for the same proper time the relative position of the 
shell when back-reaction is included is given by $R_s-R_c$;
finally, in the last column we report the ratio between the
relativistic correction and the Newtonian potential on the horizon, 
(see Eq.~(3.13)).}
\label{table1}
\end{table}
The general picture is that $R_s$ collapses slower than $R_c$,
however the difference $R_s-R_c$ computed at the proper time when 
$R_c=R_H$ is a fraction of $1/\mu$ and decreases for increasing $N$. 
Therefore, in our approach, the difference between the semi-classical
trajectory $R_s$ and $R_c$ is negligible in all physical situations.
Further, relativistic corrections associated to the neglected
term $G^2\,m^2/4\,R _s ^2$ in the equation of motion (\ref{giunzione})
amount to a deviation from $R_c$ of about $13\%$ on the horizon (see
last column in Table~\ref{table1}) which is much greater than the 
computed relative deviation of the trajectory due to a changing 
mass or to quantum fluctuations (fifth column in Table~\ref{table1}).  
Therefore one can safely approximate $R_s$ with $R_c$ down to the 
Schwarzschild radius.
\par
\begin{figure}
\centerline{
\epsfysize=250pt\epsfbox{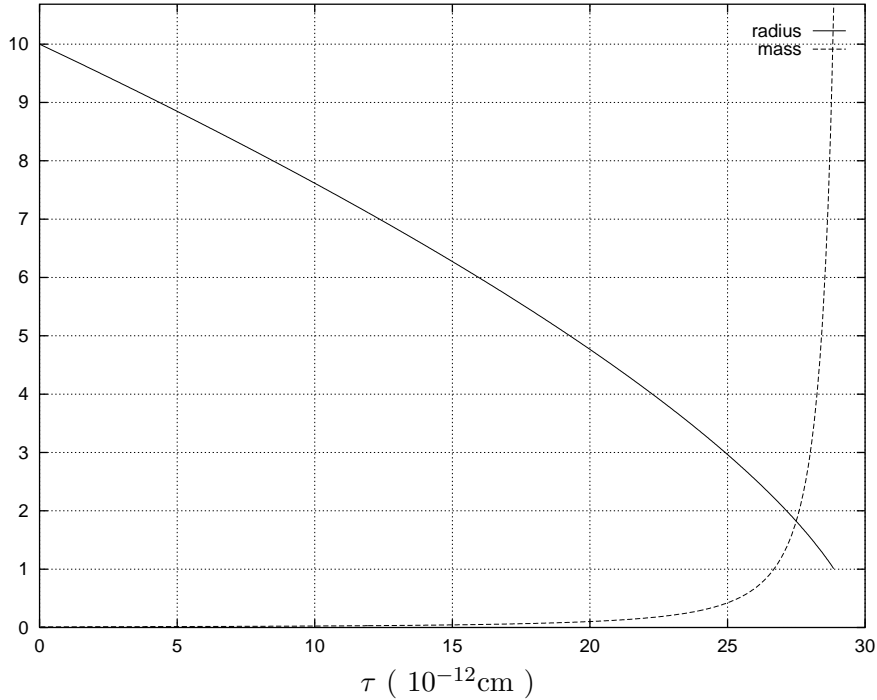}\\}
\centerline{$\tau$ ( $10^{-12}$cm )}
\caption{Evolution of the radius $R_s/R_H$ and relative mass $10^5(m-M)/M$ of 
the shell in proper time for the second case in Table~1 and $R_s(0)=10\,R_H$.}
\label{fig1}
\end{figure}

The production of matter instead is appreciable. 
In the second case displayed in Table~\ref{table1}, the number of 
produced particles $\Delta N=\Delta m/\hbar\mu$ is of order of 
$10^{-4}$ times the initial mass (see also Fig.~\ref{fig1}). 
In general the data shown in Table~\ref{table1} are in agreement
with the expected behaviour $\Delta N\propto N^{-1}$.
\par
Lastly for $R_s\to 0$ the number $\Delta N$ diverges and our approach 
breaks down completely since all the consistency conditions are violated.
\section{Conclusions}
\label{conclusion}
Our purpose has been the description of the gravitational collapse
of a self-gravitating shell of quantum matter.
We have attempted to obtain a ``realistic'' description of such
a shell by considering a bound state of a sufficiently large number
of scalar particles.
It was heuristically found that it was possible to obtain realistic
and ``thin'' shells.
\par
Since our shell consists of  scalar particles it is plausible to assume
that the particles, at least initially, will all be in the ground
state and form some sort of condensate.
Subsequent evolution during the collapse will cause transitions
to higher excited bound states which presumably, once a sufficient
number of them is formed, will decay collectively through the creation
of an additional particle in the ground state.
This will lead to a condensate with an increasing number of particles.
We have pointed out that this is somewhat analogous to the hadronization
process in QCD.
\par
The above approach then led to an effective 
time dependent Lagrangian for matter.
On including the shell Lagrangian and quantizing the matter-gravity
system we obtained the corresponding Wheeler-DeWitt equation.
On subsequently considering the semiclassical limit for gravity and
neglecting fluctuations, we introduce a matter Fock space and reproduce
the precedingly obtained Einstein equation for a many scalar particle 
shell together with the Schr\"odinger equation for matter.
The latter system was then solved using the method of adiabatic invariants.
\par
On analyzing our results, the main conclusion is that, for a suitable number 
of initial particles constituting the shell, the matter proper energy 
increases appreciably when the shell radius approaches the horizon. 
This corresponds to an increase of the number of particles, that is matter
creation.
Correspondingly, one obtains corrections to the equation of motion for
the mean shell radius which lead to a slow down of the collapse.
However, for the range of parameter considered, this change does not produce
observable effects on the trajectory of the shell. 
It is clear that suitable conditions should lead to more significant effects,
however our approach would then break down.
\par
It would be of interest, within our quantitative approach, to consider
an extension to many shells leading to a more general matter distribution 
in order to study its collapse, again taking into account matter quantum 
effects.
We hope to return to this point.
%
%
%
%


\begin{thebibliography}{99}
\bibitem{asy}
We assume the space-time is asymptotically flat and no sources other
than the shell itself are present.
\bibitem{again}
Again one has an analogy with hadronic physics in which an excited 
hadron (long string) decays into lighter ones (short strings).
\end{thebibliography}
\end{document}